# Survey of Exciton-Phonon Sidebands by Magneto-optical Spectroscopy Using Highly Purified (6,5) Single-walled Carbon Nanotubes

Weihang Zhou,[1] Tatsuya Sasaki,[1,4] Daisuke Nakamura,[1] Hiroaki Saito,[1,4] Huaping Liu,[2,3] Hiromichi Kataura,[2,3] Shojiro Takeyama[1,*]

[1]*The Institute for Solid State Physics, The University of Tokyo, 5-1-5, Kashiwanoha, Kashiwa, Chiba 277-8581, Japan*

[2]*Nanosystem Research Institute, National Institute of Advanced Industrial Science and Technology, Tsukuba, Ibaraki 305-8562, Japan*

[3]*Japan Science and Technology Agency, CREST, Kawaguchi , Saitama 330-0012 , Japan*

[4]*Department of Applied Physics, The University of Tokyo, Hongo 113-8656, Japan*

**ABSTRACT:**

We report the first high-field magneto-optical study on the exciton-phonon sideband of single-walled carbon nanotubes (SWNTs) consisting only of (6,5) species. Both energy and intensity of the observed phonon sideband were found to be independent of the external magnetic field. Comparing with theoretical calculations, we confirmed that these sidebands originate from the optically forbidden K-momentum singlet excitons. Energy of these K-momentum dark excitons was extracted to be ~ 21.5 meV above the bright Γ-momentum singlet excitons, in close agreement with recent theoretical predictions and experimentally determined values.

Single-walled carbon nanotubes (SWNTs) are rolled-up two-dimensional graphite sheets with one atom thickness. As one of the most important nano-scale materials, optical properties of SWNTs have attracted lots of attention in recent years [1-4]. It is now generally known that optical properties of SWNTs are dominated by excitonic effects, due to the exceptionally large excitonic binding energy induced by their one-dimensionality [3, 5-6]. Because of the degeneracy of the K and K$^/$ valleys, together with electron spins, excitonic structure of SWNTs is predicted to comprise of 16 excitonic states, among which only the spin-singlet zero-momentum excitonic state is optically allowed (bright) and all others optically forbidden (dark) [7-8]. A thorough understanding of these excitonic structures, including both bright and dark states, is of vital importance for potential applications of SWNTs in future opto-electronic devices.

Recent studies show that exciton-phonon coupling plays an important role in understanding the optical spectroscopy of SWNTs [9-11]. Strong exciton-phonon coupling opens up new recombination and emission channels for dark excitonic states, providing an efficient way for the study of dark levels of SWNTs. Such exciton-phonon complexes were found to manifest themselves as absorption satellites lying ~ 200 – 220 meV above (referred to as X2) or emission satellites lying ~ 120 - 140 meV below (referred to as X1) the main exciton peak [9, 11, 12-13]. The asymmetric location of X1 and X2 about the bright Γ-momentum singlet exciton (referred to as B-Γ-S exciton) leads to the suggestion that X1 and X2 are phonon sidebands of the dark K-momentum singlet exciton (referred to as D-K-S exciton) coupled with the zone-boundary $A_1$' symmetry phonons (henceforth referred to as Model #1) [14-18]. However, different models have been proposed by other studies, in which X2 was ascribed to be phonon sidebands of the B-Γ-S exciton coupled with zone-center $\Gamma_{LO}$ phonons, and X1 assigned to exciton peak of some dark deep excitonic state lying below the bright $E_{11}$ exciton (henceforth referred to as Model #2) [13, 19].

On the other hand, it is now well-known that an external magnetic field applied parallel to the tube axis changes the electronic structure and thus the excitonic states of SWNTs drastically [7-8, 20-23]. The field dependent excitonic structure of SWNTs provides an ideal test-bed for the study of Aharonov-Bohm effect in realistic systems. Moreover, what should be noted is that the evolution of the phonon sidebands in magnetic fields is correlated to their parent excitonic states. Thus, the distinctive evolution of excitonic states in external fields can be used as an efficient tool to clarify the physical origin of the observed phonon sidebands.

Following this idea, we prepared stretch-aligned SWNTs which essentially contain (6,5) species only and performed high-field magneto-optical absorption measurements, with the aim to add clarity to the recent debate about the origin of phonon sidebands. We also note that magnetic field effect on SWNT phonon sidebands remains mysterious so far, although phonon-mediated process plays an essential role in understanding the optical properties of SWNTs. To the best of our knowledge, this is the first experimental study on the magnetic field dependent behaviors of SWNT phonon sidebands.

The highly purified (6,5)-only SWNTs were isolated from a high-pressure carbon monoxide (HiPco)-grown mixture using the single-surfactant multicolumn gel chromatography method [24]. For the alignment of SWNTs, we first mixed the (6,5) SWNT dispersion with polyvinyl alcohol (PVA) solution and dried the mixture in the air. Then, the SWNTs / PVA film was stretched to ~ 5 times its original length by a home-made automated stretching machine. Typical stretched SWNTs / PVA film was shown in the inset of Figure 1. For the magneto-optical absorption measurements, we used a pulsed halogen lamp as the light source. Transmitted light from the SWNTs / PVA film was guided by optical fibers and dispersed by a monochromator (Acton, SpectraPro 2300i) with focal length of 300 mm and finally recorded by a liquid Nitrogen-cooled InGaAs detector. Pulsed magnetic field was generated by a home-made wire-wound solenoid coil with a maximum field of ~ 52 T and duration time of ~ 40 ms. All magneto-absorption spectra was taken on the top of the pulsed field at liquid Nitrogen temperature, under the Voigt geometry.

Figure 1 shows the absorption spectrum of the highly purified (6,5) SWNT dispersion taken at room temperature. Series of sharp absorption peaks can be seen clearly from the figure. Comparing with theoretical calculations and reported data from others [25-26], we were able to assign most of these peaks to the (6,5) SWNT species, as labeled in the figure. The main $E_{ii}$ (i – subband index) bright exciton peaks can be identified from $E_{11}$ up to $E_{44}$ unambiguously, showing high quality of the sample. Besides these main interband transitions, several relatively weak peaks appear in the spectrum. The phonon sideband, which is under intense debate these years, is also clearly observed to be located ~ 205 meV above the $E_{11}$ main peak and is marked with "X2" in the spectrum. The one marked with "S", which lies ~ 215 meV above the $E_{22}$ peak, is identified to be phonon sideband of the $E_{22}$ manifold. The one marked with an asterisk probably comes from another species. The vanishing absorption intensity of the other minor species clearly demonstrates the exceptionally high purity of the (6,5) SWNTs.

Due to their one-dimensionality, the optical responses of SWNTs are highly anisotropic. Excitation by light polarized perpendicular to the tube axis is strongly suppressed because of the large depolarization effect [27]. Based on this unique property, the nematic order parameter, whose definition is $S = (3 <\cos^2\theta> - 1) / 2$ (θ: average angle between the SWNT axis and the selected polarization direction of the incident light.), could be deduced. Value of the average angle θ can be extracted by correlating $S$ to the optical anisotropy of SWNTs, defined as $A = (\alpha_{//} - \alpha_{\perp}) / (\alpha_{//} + 2\alpha_{\perp})$, where $\alpha_{//}$ and $\alpha_{\perp}$ are the absorption intensities for light polarized parallel and perpendicular to the SWNT axis, respectively [22]. To evaluate the alignment of SWNTs in our stretch-aligned samples, we measured the anisotropy of optical absorption in the $E_{11}$ energy region, as shown in the inset of Figure 1. It can be seen clearly that absorption of light polarized perpendicular to the SWNT axis is strongly suppressed compared with the parallel case, implying excellent alignment of SWNTs inside the PVA film. From these anisotropic spectra, we deduced the nematic order parameter $S \approx 0.63$, giving rise to an average angle $<\theta> \approx 30^0$.

Having confirmed the high quality and alignment of the stretch-aligned SWNTs / PVA film, we continued to perform magneto-absorption measurements with the aim of clarifying the origin of the exciton-phonon sidebands. Typical spectra taken at liquid Nitrogen temperature under configuration of B // E // stretching direction are shown in Figure 2. Each spectrum has been carefully calibrated and slightly offset to have the same baseline level. Two absorption peaks, corresponding to $E_{11}$ bright exciton and phonon sideband X2, are clearly observed at each field. It is now well known that an external magnetic field parallel to the SWNT axis changes the band structure drastically via the Aharonov-Bohm (AB) effect [7-8, 20-23]. The applied flux introduced a phase factor to the electron wave function and lifts the degeneracy between the $K$ and $K'$ states, eventually resulting in two independent bright excitonic states corresponding to $KK$ and $K'K'$ excitons. With increasing magnetic field, the $E_{11}$ main peak shows rapid decrease in the peak height and obvious peak broadening, as a result of the AB splitting. However, the phonon sideband X2 at different fields overlaps with each other completely, in strong contrast with the $E_{11}$ main peak.

To analyze the physical origin of the phonon sidebands, we first extracted the evolution of the $E_{11}$ main peak. The absorption spectra were deconvoluted into the Γ-momentum bright and dark excitonic components by means of fitting using the exponentially modified Gaussian (EMG) waveform [28-29]. The results of the fitting for the peak position and intensity at different magnetic fields were shown in Figure 3 (a) and (b), respectively. Previous studies show that the AB splitting of the peaks can be described well by the following formulae [22, 30]:

$$\varepsilon_{\beta,\delta}(B) = E_g \pm \frac{\sqrt{\Delta_{bd}^2 + \Delta_{AB}^2(B)}}{2} \quad \text{(peak energy)}$$

$$I_{\beta,\delta}(B) = \frac{1}{2} \pm \frac{1}{2}\frac{\Delta_{bd}}{\sqrt{\Delta_{bd}^2 + \Delta_{AB}^2(B)}} \quad \text{(peak intensity)}$$

Here, $\Delta_{bd}$ represents the zero-field splitting between the bright (β) and dark (δ) excitonic components. And $\Delta_{AB} = \mu B \cos\theta$ is the so-called AB splitting, with θ being the average angle between the SWNT axis and the applied magnetic field. Fitting of the above formulae to the peak energy and intensity evolutions gives $E_g$ = 1.235 eV, μ = 0.57 meV / T and $\Delta_{bd}$ = 8.8 meV. Values of these fitting parameters are further confirmed by our high field measurements up to 190 T on the same sample, which gives $E_g$ = 1.237 eV, μ = 0.57 meV / T and $\Delta_{bd}$ = 8.5 meV [31].

Having extracted the detailed parameters for the magnetic field evolution of the $E_{11}$ main peak, we now turn to the magnetic field dependent behavior of the X2 phonon sidebands. In the proposed physical Model #1, X2 was ascribed to be phonon sidebands of the D-K-S excitonic state (|KK' & K'K>) coupling with the $A_1$' symmetry, in-plane transverse optical phonon (iTO)

whose momentum is at the K (K') point. The dark singlet state |KK' & K'K⟩ has center of mass momentum near the K (K') point of the graphene Brillouin zone, and thus scattering of the zone-boundary $K_{A1'}$ phonon conserves momentum in the optical process. Theoretical calculations show that energy of this D-K-S exciton does not change with magnetic field within the available field range [8]. Moreover, energy of the zone-boundary $K_{A1'}$ phonon also remains constant within available fields. Thus, energy of phonon sideband X2, $E_{X2} = E_K + E_{A1'}$, remains unchanged in external magnetic fields, if its origin follows that proposed by Model #1. Energy evolutions of the D-K-S exciton and phonon sideband in the case of Model #1 were plotted by black solid and dash-dotted lines in Figure 4(a), respectively. In Model #2, however, X2 was assigned to be phonon sidebands originating from coupling of the B-Γ-S exciton with zone-center $\Gamma_{LO}$ phonons (G band). The energy shift of the B-Γ-S exciton has already been deduced from the fitting of the AB splitting, as shown in Figure 3(a). The magnetic field dependent energy shift of the phonon sideband, in proposed Model #2, can be calculated with formula $E_{X2} = E_{11}(Bright) + E_{LO}$. Using deduced parameters for $E_{11}$ bright exciton and reported energy of $\Gamma_{LO}$ phonon (~197 meV, Ref [18]), the peak shifts of the $E_{11}$ bright exciton and the phonon sideband in external fields were plotted in Figure 4(a) as red solid and dashed lines, respectively.

It has been shown recently that the line shape of X2 phonon sideband originates from the convolution of the exciton and phonon densities of states [14,15,18]. What is important is that, the actual peak position, $E_{X2}$, is shifted to higher energy because of the convolution. To extract the actual peak energy precisely, we followed the procedure employed in Refs [14,15,18] and utilized in our fitting the exponentially modified Gaussian (EMG) function, whose waveform is as follow [28-29]:

$$EMG(x) = A\exp[\frac{1}{2}(\frac{w}{C})^2 - \frac{x-x_C}{C}] \cdot \{erf[\frac{1}{\sqrt{2}}(\frac{x_C}{w} + \frac{w}{C})] + erf[\frac{1}{\sqrt{2}}(\frac{x-x_C}{w} - \frac{w}{C})]\}$$

where $A$ is the amplitude, $w$ is peak width, $x_C$ is peak center, $C$ is a constant and "$erf$" represents the error function:

$$erf(z) = \frac{2}{\sqrt{\pi}} \int_0^z \exp(-t^2)dt$$

The EMG function gives a very good fitting to the asymmetric line shape of the phonon sideband. Typical fitting result was shown in Figure 3(c). The peak energy of the X2 phonon sideband extracted from the EMG fitting is plotted in Figure 4(a) (blue rhombus). It could be seen clearly from the figure that the energy of X2 does not change at all even up to the maximum field of ~ 52 T within experimental errors. Compared with predictions of Model #1 and Model #2, we can see clearly that the magnetic field dependent behavior of X2 follows the predictions of Model #1 exactly, giving strong support to the interpretation that X2 is phonon sideband of the D-K-S exciton coupled with the zone-boundary $K_{A1'}$ phonons. From these EMG fittings, we deduced that the actual peak position of X2 (pink vertical line in Fig. 3(c)) is shifted by ~ 19.6 meV from the peaks observed in the magneto-absorption spectra, which is in good agreement with the value given by Ref [18] and slightly larger than the value of 11 meV from Ref [14].

Besides the magnetic field dependent behavior of the phonon sideband energy, we note that origin of the phonon sidebands can also be inferred from the evolution of their intensities in external fields. Without magnetic fields, the Γ-momentum singlet state |KK-K'K' (+)⟩ is the only optically active state. When an external field is applied, magnetic flux passing through the nanotube brightens the otherwise dark antibonding state |KK-K'K' (-)⟩, due to the additional AB phase introduced by the magnetic field [7-8, 20-23]. The antibonding state |KK-K'K' (-)⟩ gains oscillator strength gradually at the expense of the bright excitonic state |KK-K'K' (+)⟩, as has been shown in Figure 3(b). On the assumption that magnetic field-induced changes of the coupling strength between excitons and phonons are negligibly small, the intensity evolution of the phonon sideband should follow that of its parent excitonic state. Based on this assumption, we calculated the intensity evolution of the phonon sideband for Model #2, which ascribes X2 as phonon sidebands of the B-Γ-S excitons coupled with zone-center $Γ_{LO}$ phonons, using parameters extracted from the AB fitting. The predicted curve is shown as red dashed line in Figure 4(b), in which the sideband intensities have been normalized to their initial intensities at zero-fields, respectively. In the case of Model #1, the D-K-S exciton involved behaves in a completely different way. Theoretical studies show that the D-K-S excitonic state remains optically inactive even in external magnetic fields [8]. On the assumption of negligible changes for exciton-phonon coupling strength, intensity of phonon sideband generated from the D-K-S exciton remains the same with and without magnetic fields. Predicted result has been shown as black dash-dotted line in Figure 4(b). The experimental intensity of X2 extracted from the EMG fitting is shown as pink pentagon in Figure 4(b). Again, the intensity has been normalized to its initial intensity at zero-field, for the convenience of an explicit comparison with predictions. From the figure, it is unambiguously demonstrated that, within experimental errors, intensity of the phonon sideband does not change with magnetic field. This behavior follows exactly the prediction of proposed Model #1, in sharp contrast to the prediction of Model #2. Analysis of the intensity evolution again gives strong support to the interpretation that X2 is phonon sideband of the D-K-S exciton coupled with zone-boundary $K_{A_1'}$ phonons.

Energy of these D-K-S excitons is also a subject of wide interest in recent years [14,15]. Studies on these dark states have been difficult since they cannot couple with light directly due to their large center of mass momenta. Our analyses on the energy and intensity evolutions of X2 confirm that it is phonon sideband of the D-K-S exciton coupled with zone-boundary $K_{A_1'}$ phonon. Knowing this, we can deduce the energy of the D-K-S exciton directly:

$$E_K = E_{X2} - E_{A_1'}$$

Our EMG fitting gives an average value of X2 as $E_{X2}$ = 1424.4 meV. Using the value of $A_1'$ phonon energy from Ref [18] $E_{A_1'}$ = 163 meV, we obtained the energy of the D-K-S exciton $E_K$ = 1261.4 meV. From the peak position of $E_{11}$ bright exciton $E_{11}$ = 1239.9 meV, we deduced the energy splitting between the D-K-S exciton and the B-Γ-S exciton: $\Delta_K = E_K - E_{11}$ = 21.5 meV. This calculated value is in closed agreement with the theoretical value of 23 meV predicted for SWNTs having the same diameter with the (6,5) species [32]. Furthermore, our value compares well with the experimental values of 25 meV and 29 meV reported recently [14,18]. Consistency between these data clearly shows the high experimental accuracy of our measurements.

In summary, we report the first high-field magneto-optical study on the exciton-phonon sideband of SWNTs. Single chirality (6,5) SWNTs were prepared by the single-surfactant multicolumn gel chromatography method. High degree of alignment was achieved by embedding SWNTs into PVA film and stretching it to ~ 5 times its original length. We extracted the precise value of the energy and intensity of SWNT phonon sidebands by introducing the exponentially modified Gaussian function. Independent analyses on the energy and intensity evolutions of SWNT phonon sidebands in external magnetic fields come to the same conclusion, that the observed sidebands are phonon sidebands of the K-momentum dark singlet exciton coupled with zone-boundary $K_{A_1'}$ phonons. In addition, we deduced that the optically inactive K-momentum singlet exciton lies ~ 21.5 meV above the bright Γ-momentum exciton, in closed agreement with recent theoretical prediction and experimentally determined values. Our results thus add clarity to the recent debate about the origin of SWNT phonon sidebands and bring people a deeper understanding on the electronic structure of SWNTs. We hope these results could be helpful for future experiments and potential applications of SWNTs in opto-electronics.

We are obliged to Prof. K. Kindo for supplying a nondestructive pulsed magnet. One of the authors (W. H. ZHOU) thanks for the financial support of the post-doctoral research fellowship at the Institute for Solid State Physics.

# Figures

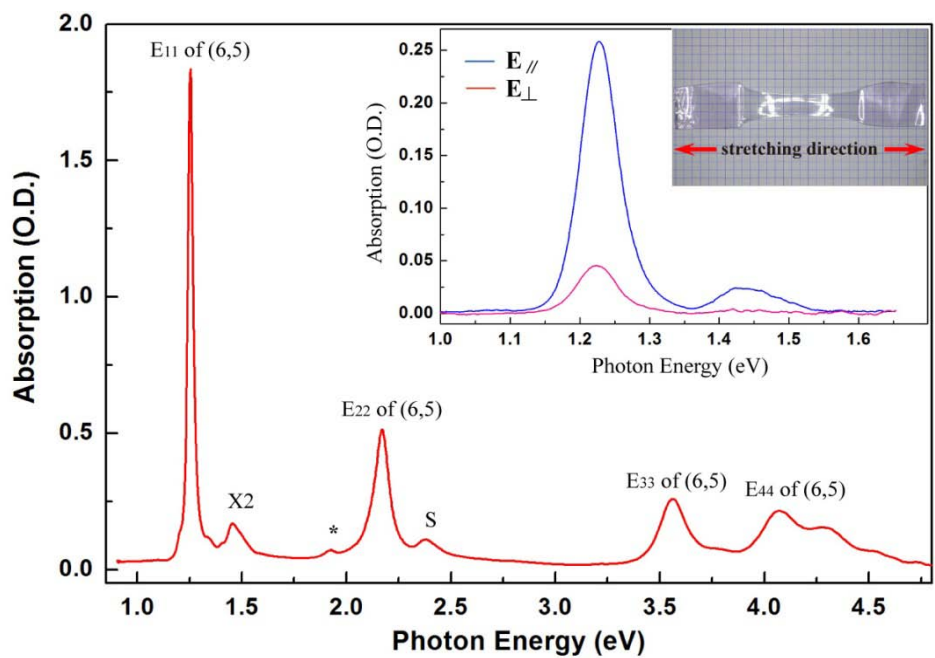

Figure 1. Absorption spectrum of highly purified (6,5) SWNTs dispersed in $D_2O$. Inset: Absorption spectra of stretch-aligned (6,5) single chirality SWNTs / PVA film for light polarized along ($E_{//}$) and perpendicular ($E_\perp$) to SWNT axis. Together shown in the inset is photo of the stretched SWNTs / PVA film. Measurements were taken at room temperature.

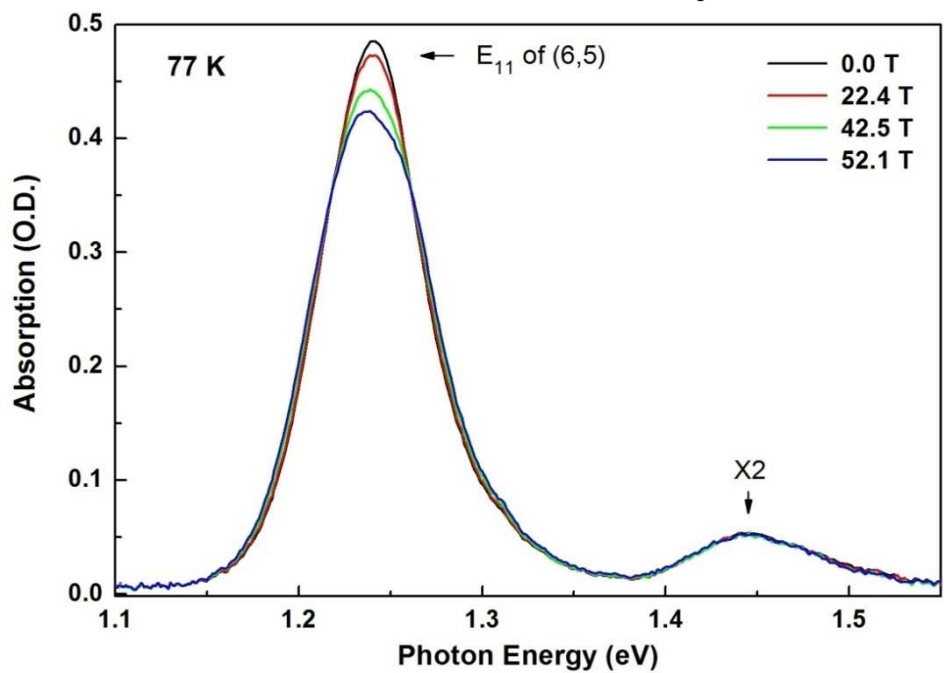

Figure 2. Typical magneto-absorption spectra of stretch-aligned (6,5) SWNTs / PVA film sample taken at 0.0 T, 22.4 T, 42.5 T and 52.1 T, respectively. Measurements were taken at liquid Nitrogen temperature with field direction and light polarization parallel to the stretching direction of SWNTs / PVA film (B // E // Stretching direction).

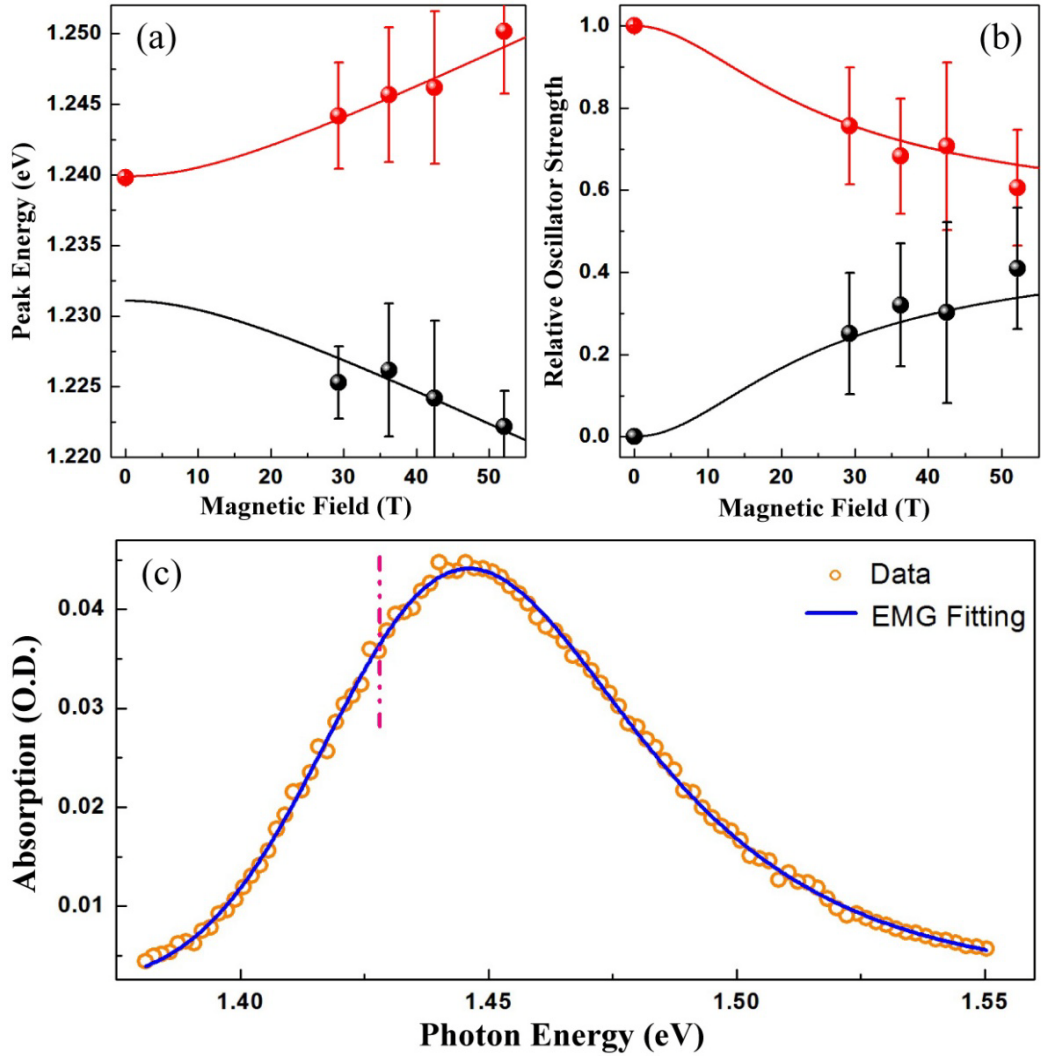

Figure 3. (a) Energy of bright and dark excitonic components as a function of external magnetic fields. (b) Intensity of bright and dark excitonic components as a function of magnetic field. Red filled circle: bright exciton; Black filled circle: dark exciton; Red and black solid lines: fitting curves. (c) Phonon sideband absorption spectrum at 0.0 T (orange circle) and its best-fit curve using the EMG function (blue solid line). Pink vertical line denotes the actual peak position extracted from the EMG fitting.

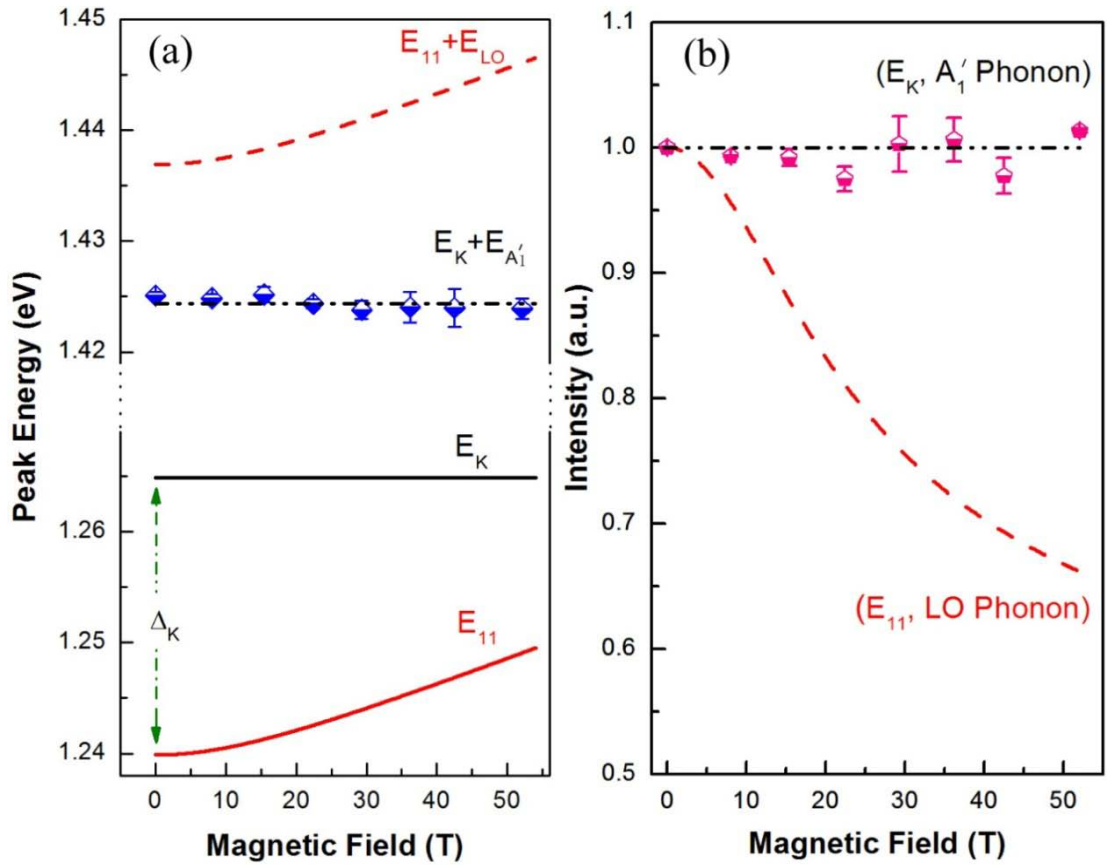

Figure 4. (a) Peak energy of phonon sideband (X2) as a function of magnetic fields and theoretically predicted curves. Blue rhombus: experimental data; Red solid line: energy of the B-Γ-S exciton calculated with parameters: $E_g$ = 1.235 eV, μ = 0.57 meV/T, $\Delta_{bd}$ = 8.8 meV; Red dashed line: predicted energy of the phonon sideband in the case of Model #2; Black solid line: energy of the D-K-S exciton; Black dash-dotted line: predicted energy of the phonon sideband in the case of Model #1. (b) Intensity of phonon sideband as a function of magnetic field and theoretical predictions. Pink pentagon: experimental data; Red dashed line: intensity of phonon sideband in the case of Model #2; Black dashed dotted line: intensity of phonon sideband in the case of Model #1. All the intensities have been normalized to their zero-field intensities, respectively.